\documentclass{PoS}
\usepackage{booktabs}
\usepackage{url}
\usepackage{appendix}

\title{Future colliders - Linear and circular}

\ShortTitle{Future colliders}

\author{\speaker{Roman P\"oschl}
        Laboratoire de l'Acc\'el\'erateur Lin\'eaire (LAL)\\
        Centre Scientifique d'Orsay, B\^at. 200\\
        F-91898 Orsay\\
        E-mail: \email{poeschl@lal.in2p3.fr}}

\abstract{The completion of the Standard Model of particle physics by the discovery of a light Higgs boson at the LHC in 2012 triggered the debate about the best way forward to discover physics beyond the Standard Model. At the eve of the update of the European Strategy of particle physics, this article summarises the motivation and status of the different collider projects in particle physics at the energy frontier. This article reflects the status as of December 2018 and the reader is invited to follow closely updated information of the presented projects.}

\FullConference{An Alpine LHC Physics Summit (ALPS2018)\\
		15-20 April, 2018\\
		Obergurgl, Austria}

\begin{document}

\section{Introduction}

The discovery of a Higgs boson and the absence of new physics signals at the LHC shape the strategy on future projects in particle physics. As of today several projects with different levels of maturity are under discussion. The projects can be distinguished between hadron, electron-positron colliders and electron-hadron colliders. Namely, these projects are:
\begin{itemize} 
\item Hadron colliders
\begin{itemize} 
\item High luminosity LHC (HL-LHC);
\item High energy LHC (HE-LHC);
\item Future Circular Collider for hadrons (FCC-hh);
\item Super Proton Proton Collider (SppC).
\end{itemize}
\item Electron-positron colliders
\begin{itemize} 
\item International Linear Collider (ILC);
\item Compact Linear Collider (CLIC);
\item Future Circular Collider for electrons and positrons (FCC-ee);
\item Chinese Electron-Positron Collider (CEPC).
\end{itemize}
\item  Electron-hadron colliders
\begin{itemize} 
\item HL-LHeC, HE-LHeC and FCC-eh.
\end{itemize}

\end{itemize}

In the following a brief review of the physics landscape will be given and the machines will be briefly described. For further details the reader may consult the references given in the text.

\section{Physics landscape}

The ongoing analysis of the data recorded at the LHC shows that the scalar particle discovered in 2012 agrees with predictions of the Standard Model of particle physics for the Higgs boson. This Higgs boson has a mass of $125.16\pm0.16$\,GeV~\cite{bib:pdg2018}. Meanwhile the decay of the Higgs boson into pairs of bottom quarks~\cite{ATLAS-CONF-2018-036, Sirunyan:2017guj} and $\tau$ leptons~\cite{Aad:2015vsa,Sirunyan:2017khh} have been observed as well as recently the radiation off a top quark in $\mathrm{t\bar{t}H}$ production at the LHC~\cite{Aaboud:2018urx, Sirunyan:2018hoz}. The small mass of the Higgs boson and sizeable corrections to the mass when taking into account higher orders in perturbation theory require, within the Standard Model, an unsatisfactory fine tuning between the bare and the renormalised mass. Further, the Standard Model does not contain a dark matter candidate whereas the existence of dark matter is suggested by cosmological observations. Popular theories to remedy these shortcomings as Supersymmetry fail to be discovered at the LHC. The exclusion limits for coloured supersymmetric particles is currently of the order of 1\,TeV. Alternatives to Supersymmetry are composite models or models with extra-dimensions. Today's understanding is that the direct observation of the particle content of these models is beyond the reach of the LHC or that the precision that can be achieved at the LHC is not sufficient for indirect discovery. There are in general two approaches to make progress: 
\begin{itemize}     
\item Increase the achievable precision of the measurements of the discovered Higgs boson and of particles that carry the {\it imprint of the Higgs particle} such as $W$ and $Z$ bosons or the top quark;
\item Operation at higher centre-of-mass energies to access first mass states of new physics. 
\end{itemize}

\section{Current and future hadron colliders}

Hadrons are composed objects with the total energy carried by the particle shared in a probabilistic way among its constituents. Therefore hadron colliders allow for sweeping over a large energy range exploring thus a large phase space for discoveries. The most powerful hadron collider today is the LHC that collides protons with a centre-of-mass energy of 13\,TeV (after an initial phase with a centre-of-mass energy of 8\,TeV). The upgrade of the machine to the so-called high-luminosity LHC (HL-LHC) has been approved and is currently being prepared~\cite{Apollinari:2284929}. The typical instantaneous luminosity produced by LHC in 2018 is $2\times 10^{34}\,\mathrm {cm^2 s^{-1}}$.  For the high luminosity phase the targeted value will be about $5\times 10^{34}\,\mathrm {cm^2 s^{-1}}$ at 14\,TeV. 
The increase of luminosity will be achieved by an increase of the beam current by about a factor of two. At the same time the beam emittance will be reduced by one third. This requires innovative technologies that include high field, i.e. 11 T, dipole magnets and a crab cavity scheme to provide a small $\beta^{\ast}$ at the interaction point. 
The high luminosity comes along with so-called pile-up events. Beyond the interesting `hard` interaction there will be also up to 135 parasitic soft interactions that need to be mitigated by the experiments.

\subsection{HE-LHC, FCC-hh and SppC} 

Following a recommendation of the update of the European Strategy of Particle Physics in 2013~\cite{bib:2013kh}, CERN in collaboration with international partners is conducting an R\&D program for the next generation hadron machines. This effort is complemented by a similar program initiated by the Chinese Particle Physics Community. A summary of parameters of future hadron colliders is given in Table~\ref{tab:hhparams} in Appendix~\ref{app:a}.

The intuitive next step is to reuse the existing LHC tunnel and to install a machine that reaches an about a factor of two higher centre-of-mass energy than the LHC or the HL-LHC. Targeted are energies of $\sqrt{s}\approx 27\,\mathrm{TeV}$. This proposal is known as high energy or HE-LHC. To go significantly beyond that number a new infrastructure is needed. The FCC-hh project proposes a machine with a centre-of-mass energy of $\sqrt{s}\approx 100\,\mathrm{TeV}$ that would reside in a circular tunnel of about 100\,km in circumference. The Chinese SppC plans equally for a tunnel of 100\,km at a designed centre-of-mass energy of  $\sqrt{s}\approx 75\,\mathrm{TeV}$~\cite{CEPCStudyGroup:2018rmc}. The HE-LHC and the FCC-hh will present a Conceptual Design Report as input to the next update of the European Strategy of Particle Physics to be concluded in 2020.

\begin{figure}[ht]
\centering
\includegraphics[width=0.99\textwidth]{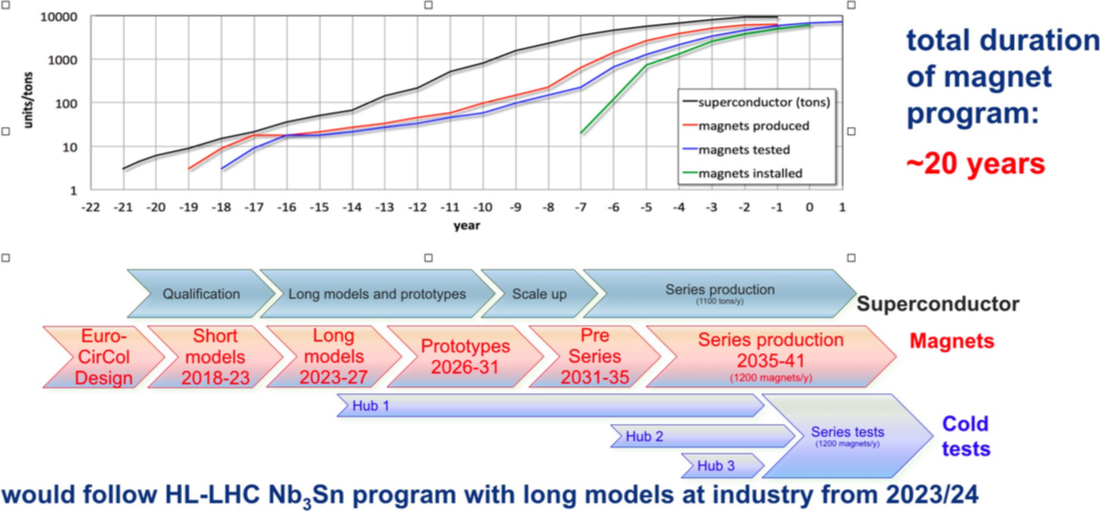}
\caption{\label{fig:16T-magnet} \sl Overview on the R\&D path towards 16T dipole magnets for future $pp$ colliders. The figure is taken from~\cite{bib:mangano-ias2018}.}
\end{figure}

The biggest technological challenge of these proposals is the realisation of high-field dipole magnets. For the HE-LHC and the FCC-hh these magnets will have to provide a magnetic field of 16 T over a length of about 13.5\,m (the final value depends on the beam optics)~\cite{bib:angeles-jtlal2018}. The SPPC envisages values between 12\,T (initial design) and 24\,T at a length of 15\,m~\cite{CEPCStudyGroup:2018rmc}. At the highest values the targeted current densities are of the order of $1500\,\mathrm{A/mm^2}$. The development of these magnets is part of a worldwide program for innovative magnet technologies. This implies in particular the application of $\mathrm{Nb_{3}Sn}$ as conductor material that allows for higher current densities as the commonly used $\mathrm{NbTi}$ Conductors. These conductors will be used already in the HL-LHC for quadrupole magnets and also for a few 11 T dipole magnets that are going to be installed in HL-LHC. As shown in Fig.~\ref{fig:16T-magnet} the entire development plan towards the installation of a large quantity of high-field dipole magnets in particle physics accelerators extends over about 20 years.  

\section{Future electron-positron colliders}

The impressive establishment of the electroweak theory as being correct up to the so far accessible
energies is based on a tight interplay between hadron accelerators on one hand (SppS, Tevatron) and electron-positron colliders
on the other hand (PETRA, TRISTAN, LEP, SLC). The precision of hadron machines is limited by the complicated structure of the initial state hadrons and the huge background from QCD induced reactions. Deep understanding of the physics beyond the Standard Model requires thus the parallel running of a precision machine. 

This is even more true after the discovery of a Higgs boson in 2012. In order to cover the full Standard Model phenomenology centre-of-mass energies between the $Z$-Pole and about 1 TeV are desirable. As already proven by SLC, polarised beams are an essential asset for precision tests of a chiral theory such as the Standard Model. Circular colliders as e.g. LEP are heavily limited by the energy loss by synchrotron radiation (SR energy loss).  The energy loss per turn of a particle with mass $m$ and an energy $E$
can be expressed as
\begin{equation}
\frac{\Delta E}{{\rm Turn}} \sim \frac{\gamma^4}{r}\,\,\,\,\mathrm{with}\,\gamma=\frac{E}{m}
\end{equation}
and $r$ being the radius of the circular accelerator or storage ring.

Therefore the R\&D in the last around 20 years has been concentrated on the development of linear accelerators. At least one of these projects, the ILC, has reached the necessary level of maturity to be proposed as the next major project of particle physics. An alternative solution that will be debated in the European Strategy discussion is the CLIC project. Current plannings comprise however also proposals for circular electron-positrons accelerators. This article will put some emphasis on the ILC but all alternatives will be discussed in the following. A summary of parameters for variants of the machines that are likely to become the respectively first stage is given in Table~\ref{tab:eeparams} in Appendix~\ref{app:b}.

\subsection{The International Linear Collider - ILC}
Following an international evaluation process~\cite{bib:itrp2004}, it turned out that a technology based on superconducting accelerating cavities is best suited to meet the goals of a machine that operates between the $Z$-Pole and around 1\,TeV. This technology is at the basis of the ILC concept. The ILC is designed to be operated with 9-cell niobium cavities cooled by super-fluid Helium to 2\,K~\cite{bib:tesla-tdr, bib:rdr2007}. 
In comparison with other concepts based on normal conducting cavities two major advantages can be highlighted.
\begin{itemize}
\item The ILC cavities have comparatively high quality factor of typically $10^{10}$. Therefore the power transfer of the accelerating radio frequency wave to the beam is very high. About 10\% of the overall wall-plug power, $P_{AC}$, is transferred to the beam. 
\item The operation frequency of superconducting cavities is limited by the surface resistance which increases with increasing frequency.
Since the eigen frequency of a resonator is inversely proportional to its diameter, the cavities have to be comparatively large.
However, the large size of the superconducting cavities has the advantage that wake field effects, i.e.\,a mutual influence 
of neighboured bunches due to their proper electromagnetic fields, are reduced. Wake field effects occur when particles are displaced transversally from their nominal orbit. This reduced sensitivity of the accelerating structure with respect to wake field effects allows also for somewhat relaxed tolerances in the alignment of the cavities that in turn allows for an easier operation of the accelerator.

\end{itemize}

Polarised electrons are produced by means of a Ti:sapphire laser impinging on a GaAS diode. The degree of polarisation at the interaction point will be around 80\% (90\% at the source). The baseline for polarised positron production is a helical undulator. Here, the polarisation of photons is transferred to electron-positron pairs in a dedicated target and the positrons are extracted. The expected degree of polarisation ranges between 20\% - 30\% depending on the centre-of-mass energy. Note for completeness that a conventional unpolarised positron source is also being considered. 

The ILC has presented a Technical Design Report in 2013~\cite{Behnke:2013xla}. This TDR assumes 500\,GeV centre-of-mass energy as the baseline. After the TDR the R\&D continued with the aim to reduce the cost of the machine. Among others, this effort resulted in a revision of the surface treatment of the superconducting cavities. This leads to an increase of the gradient by about 10\%. This allows for expecting the average gradient to be 35,MV/m compared with 31.5\,MV/m as in the TDR. Thus the machine can become shorter at the same energy yield. Further the Q-factor has been improved  by a factor of two ($0.8\times10^{10}\rightarrow 1.6\times 10^{10}$) yielding a higher efficiency in the cryo-system and thus a more cost effective operation.

\begin{figure}[ht]
\centering
\includegraphics[width=0.99\textwidth]{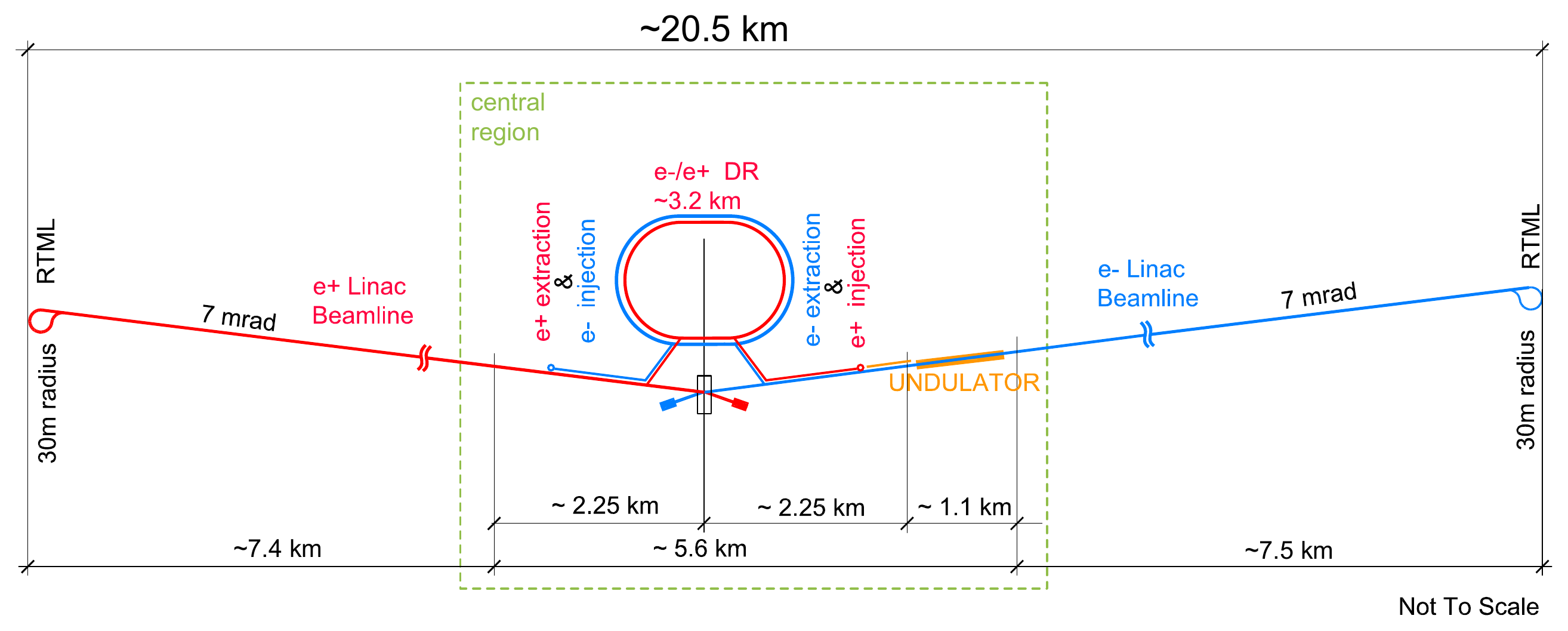}
\caption{\label{fig:ilc250} \sl Schematic layout of the ILC in the 250 GeV staged option. The figure is taken from~\cite{bib:lcc-ilcesu2018}}
\end{figure}

In particular the relatively small mass of the Higgs-boson implies that there is already an extremely strong physics case at or slightly above the $ZH$-threshold. A first stage of the ILC at 250\,GeV centre-of-mass energy, called ILC250 hereafter, results in a cost reduction by about 40\% w.r.t. to the baseline design~\cite{Evans:2017rvt}. 
A sketch of the ILC250 stage is shown in Fig.~\ref{fig:ilc250}. This option is currently evaluated at ministry level in Japan. It is expected that the Japanese government will express its attitude towards the project at the beginning of 2019. Such a statement would initiate negotiations with other countries with the aim to realise the ILC250 in Japan during the next decade.   

The technology proposed for the ILC has been applied for the construction of the European X-ray Free Electron Laser, E-XFEL, at DESY/Hamburg. This machine with a length of 4\,km required the production of cavities, cryomodules and the related infrastructure at industrial level. The E-XFEL has started operation in Spring 2017. Further light sources that use the ILC technology such as LCLS-II (SLAC) and SHINE at SINAP (Shanghai) are under construction. It is intuitively clear that the successful construction and operation of these machines and in particular of the E-XFEL gives a lot of credibility that also the ILC can be realised soon. Note that about 10\% of the cavities produced for the E-XFEL feature a gradient of greater than 40\,MV/m~\cite{bib:ayamam-hh2018}.  The active field of super-conducting accelerator facilities allows for expecting even higher gradients at the time of the ILC construction. 

\subsection{Compact Linear Collider - CLIC}

Higher energies than the baseline of 500\,GeV would be beneficial for Standard Model processes as Higgs self-coupling or $t\bar{t}H$. Also the W-Fusion process increases with centre-of-mass energy giving access to a significant amount of Higgs decays into a muon pair. The CLIC project has published a Conceptual Design Report in 2012~\cite{Lebrun:2012hj} that  proposes three energy stages at 0.5\,TeV, 1.4\,TeV and 3\,TeV, respectively. Higher centre-of-mass energies extend the discovery potential for e.g. purely electroweakly interacting supersymmetric particles beyond the reach of the LHC and the sensitivity to new $Z'$ bosons could be pushed towards up to 50\,TeV.  As the ILC, the CLIC will provide polarised electron beams with a degree of polarisation of 80\%. The polarised $e^-$-beam will be produced as described as above. 

The CDR has been updated in 2016~\cite{CLIC:2016zwp}. To reach the foreseen highest centre-of-mass energies a radically new accelerator concept is proposed. In brief, the accelerator complex consists of a drive beam and a main beam. The energy stored in the drive beam is transferred to the main beam to reach acceleration gradients of up to 100\,MV/m where gradients of 72\,MV/m and 100\,MV/m were taken into account for the CDR. Recently three major technical hurdles have been passed, namely 1) A high current drive beam bunched at 12 GHz, 2) Power transfer and main beam acceleration and 3) a gradient of up to 100\,MV/m in the main beam cavities. This progress has been achieved in the CLIC Test Facility CTF3. For a final project all these technologies will have to be industrialised. A first step will be the application in medium and large-scale systems. Examples are the study for a SwissFEL based on C-Band structures with a beam energy of 6\,GeV or the EU study for a compact XFEL based on X-band technology. 

In the updated  CDR an initial stage at a centre-of-mass energy of 380\,GeV, called CLIC380 herafter, is considered. The smaller centre-of-mass energy requires only one drive beam for the acceleration of both beams which naturally yields a considerable cost reduction. This version of the machine will be based on cavities with a gradient of 72\,MV/m.
At this comparatively small energy also rather traditional technologies based on high-efficiency klystrons can be considered. Note however that with increasing energy the klystron technology is going to be more expensive than the drive beam approach. The crossing point is at around 500\,GeV centre-of-mass energy of the machine.

\subsection{CEPC and FCC-ee}

The proposals of hadron machines hosted in tunnels of up to 100\,km in length together with the relatively small mass of the observed Higgs boson have motivated the idea to install, prior to the hadron machines, circular $e^+e^-$ colliders, called CEPC~\cite{bib:cepc2015, CEPCStudyGroup:2018rmc} and FCC-ee, in these tunnels. The larger radius remedies at least partially the shortcomings on circular machines due to synchrotron radiation. Circular $e^+e^-$ colliders will naturally also benefit from the progress in the acceleration gradient of cavities to compensate for the energy loss. 

The current proposals for circular machines include operations at centre-of-mass energies of 91.2\,GeV (i.e. the $Z$-pole), 160\,GeV ($WW$-threshold), 240\,GeV ($ZH$-threshold) and 365\, GeV (at or slightly above the $t\bar{t}$ threshold, the latter only for FCC-ee). The main advantage of circular colliders are higher luminosities that may be reached at energies below the $t\bar{t}$ threshold. However, to go significantly beyond what has been reached at LEP and what is possible at a linear collider, the accelerators will have to be operated at the beam-beam limit with maximal beam-beam parameter $\xi_y$ and beam beta-function parameter $\beta^{\ast}_y$. However, since $\xi_y \sim \beta^{\ast}_y$ the optimal working point will have to be identified with dedicated simulations. 
A common issue that may compromise high luminosities are beam instabilities in case of asymmetric bunch charges.  In short words the weaker beam blows up in the presence of the field of the other one. This requires `bootstrapping` meaning that both beams are filled at the same time with for example $5\times 10^9$ particles at a time. At least at high energies another effect to be observed is the beam-strahlung (i.e. mutual beam-beam interaction due to the quadrupole fields of the bunches), which reduces the instantaneous luminosity.
 
One particular issue is the short lifetimes of the beams as a consequence of the desired high luminosities. The lifetimes are between   
20 Minutes and 70 Minutes. The colliding beams will thus have to be provided by a top-up injection scheme in which particles will be accelerated e.g. in a booster ring to their nominal energy before being injected into the main ring that does not ramp up or down. The booster ring may constitute a particular challenge depending on the injection energy into that booster ring. However, the top-up injection is currently successfully applied at the SuperKEB factory. Requirements at SuperKEKB are partially even stronger that at the high-energy  $e^+e^-$ colliders. Both, the beam current and the $e^+$ production rate are higher at SuperKEKB. 

\section{Electron-hadron colliders}
The analysis of proton-proton collisions requires a profound knowledge of the inner structure of the proton. This structure is probed in electron-proton collisions as e.g. in the past by HERA. The LHeC project~\cite{AbelleiraFernandez:2012cc} proposes to add an electron accelerator to the existing LHC complex and similarly to the FCC-hh complex in case of the FCC-eh proposal. 

The electron machine in this case would be an ERL. In brief, in an ERL beam is continuously injected at a given energy. The energy stored in an injected beam is transferred to the subsequently injected one. The targeted final energy for the electron beam for HL-LHeC or HE-LHeC is 60\,GeV. This electron energy together with the energy of the LHC proton beam of 7\,TeV would increase the available phase space probing the structure function by a factor 2-3. Electron-proton collisions are also sensitive to electroweak processes.  The high centre-of-mass energies will allow for measurements of relative couplings of the Higgs boson and to study single top production.

Electron Recovery Linacs are operated at several research institutes around the world. A prototype for the ERL for LHeC is PERLE~\cite{Angal-Kalinin:2017iup}. One of the main goals of PERLE is to achieve high currents of the order of 10\,mA needed to reach the luminosity goal of O($10^{34}\,{\rm cm^{-2}s^{-1} }$) at a LHeC facility. 

\section{Conclusion and outlook} 
\begin{figure}[ht]
\centering
\includegraphics[width=0.9\textwidth]{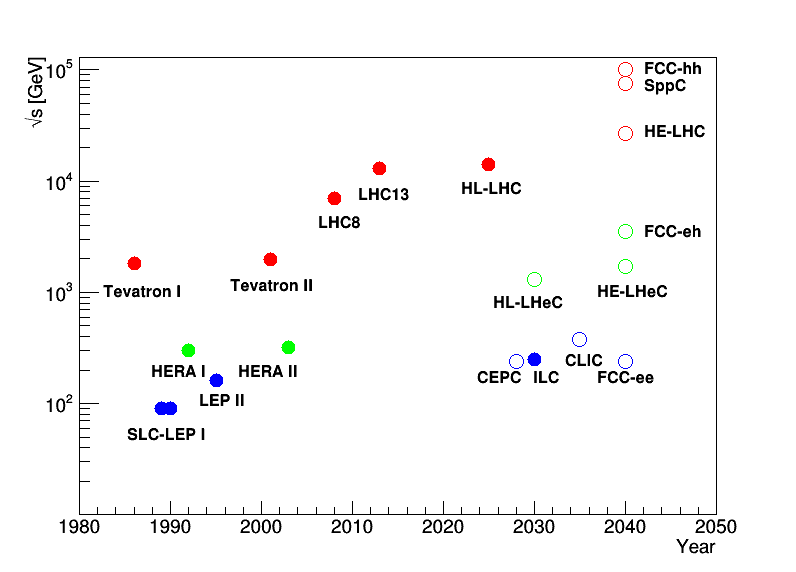}
\caption{\label{fig:accel-hist} \sl Past and present projects of $pp$ (red symbols), $ep$ (green symbols) and $e^+e^-$ (blue symbols) colliders at the energy frontier since around 1990 indicating the (tentative) start of operation and the (presumably initial) centre-of-mass energy. Full bullets represent projects that either have been completed, that are currently running or for which a TDR is a available. Open bullets represent projects that have just published, CEPC~\cite{CEPCStudyGroup:2018rmc}, or are about to publish a CDR.}
\end{figure}
As a summary Fig.~\ref{fig:accel-hist} shows past present and future projects of electron-positron,  hadron colliders and electron-hadron colliders since around 1990. All future projects require worldwide collaboration given their size and hence their cost. For the two linear colliders cost estimates of 4.8 - 5.3 BILCU\footnote{The reference currency (the `ILCU`) is the United States dollar (USD) as of January, 2012.} in case of ILC250~\cite{bib:MEXT-ilc-2018} and 6.7\,BCHF EUR in case of CLIC380 are given. For the circular electron-positron colliders Ref.~\cite{CEPCStudyGroup:2018rmc} extrapolates the cost of the LEP accelerator and arrives at 12\,BCHF if the machine would get built in Europe.  A construction in China may reduce the cost by about 50\%. The real cost may lie between these two values.
It is difficult to estimate today the full cost of a large 100\,TeV hadron collider since for example the magnet R\&D program is just at its beginning. Apart from civil construction the high-field dipole magnets are the major cost driver. Though civil facilities may be shared between circular electron-positron machines and hadron machines it seems fair to say that a e.g. FCC-hh will cost a multiple of the LHC. Obviously, a HE-LHC will cost significantly less due to its smaller size.  LHeC can be considered as an upgrade of the HL-LHC with therefore a comparable cost. 

The future of accelerator based particle physics will be shaped by decisions and strategies that will be taken or formulated in the next two years. The community waits for a statement by the Japanese government toward the realisation of the ILC. This will be taken into account in the next update of the European Strategy of particle physics. This process has been launched and will be completed in the middle of 2020.

\section*{Acknowledgements}
The author would like to thank the organisers of the ALPS2018 Conference for the invitation and the opportunity to give this talk. 
R.P. thanks Fran{\c c}ois Richard for the careful reading of the manuscript and useful comments. The Linear Collider Collaboration by virtue of Benno List and Juan Fuster have kindly put Fig.~\ref{fig:ilc250} at my disposal. Jie Gao has kindly answered promptly a trivial question on the CEPC design. Beyond the cited material the author has consulted talks by Phil Burrows and Max Klein.  
\bibliographystyle{utphys_mod}
\bibliography{alps2018}

 \begin{appendices}
\newpage

\section{Parameters of $pp$ colliders}\label{app:a}

\begin{table}[htbp]
   \centering
   \begin{tabular}{@{} ccccccc @{}} 
      \toprule
      \multicolumn{3}{c}{Parameters} \\
      \cmidrule(r){1-3} 
      Quantity    & Symbol & Unit & HL-LHC & HE-LHC & FCC-hh & SPPC\\
      \midrule
      Centre-of-mass energy  & $\sqrt{s}$ & TeV & 14& 27 & 100 & 75\\
      Number of IP &$N_{IP}$& 1 & 2 & 2 & 2 &2\\
      Luminosity/IP   & ${\cal L}$     &  $10^{34} \mathrm {cm^2 s^{-1}}$ & 5 & 15 & 5-30 & 10\\
      Background ev./bx & $N_{back}$ & 1 & 135 & 460 & 170-1000 & n.a.\\
      Arc dipole field & $B_d$ & T & 8.33 & 16 & 16 & 12-24\\ 
      Bunch population & $N_p$ & $10^{11}$ & 2.2 & 2.2 & 1 & 1.5\\
      Bunch interval   &  $\Delta t_b$ & ns & 25 & 25 & 25 & 25\\ 
      Beam current    &  $I_{beam}$ & mA & 1120 & 1120 & 500 & 730\\
      Stored beam energy  &  $E_{beam}$ & GJ & 0.7 & 1.4 & 8.4 & 9.1\\
      Normalised emitt. at IP &  $\gamma \epsilon_{x,y}$ & $\mu$m & 2.5 & 2.5 & 2.2 & 2.4\\
      Beam size@IP &  $\sigma^{\ast}_{x,y}$ & $\mu$m & 7.1 & 9.0 & 6.7-3.5 & 6.8\\
      SR power/beam & $P_{SR}$ & MW & 0.007 & 0.1 & 2.4 & 1.1\\
      Site AC Power & $P_{site}$& MW & 157 & 163 & 554 & n.a.\\
      Ring circumference & $L_{ring}$ & km & 26.7 & 26.7 & 97.8 & 100\\ 
  \bottomrule
   \end{tabular}
   \caption{Parameters of the different proposals for energy frontier $pp$ colliders (Status: December 2018). Disclaimer: The numbers have been carefully compiled or estimated from Refs.~\cite{Apollinari:2284929, bib:tang-ias2018, bib:mangano-ias2018, bib:mertens-fcc2018, bib:schulte-hh2018, Bordry:2018gri, CEPCStudyGroup:2018rmc}. All mistakes are mine and R.P. invites projects to point out mistakes or to add missing numbers.}
   \label{tab:hhparams}
\end{table}

\newpage
\section{Parameters of $e^+ e^-$ colliders}\label{app:b}

\begin{table}[htbp]
   \centering
   \begin{tabular}{@{} ccccccc @{}} 
      \toprule
      \multicolumn{3}{c}{Parameters} \\
      \cmidrule(r){1-3} 
      Quantity    & Symbol & Unit & ILC250 & CLIC380 & CEPC240 & FCC-ee ZH\\
      \midrule
      Centre-of-mass energy  & $\sqrt{s}$ & GeV & 250& 380 & 240 & 240\\
      Number of IP &$N_{IP}$& 1 & 1 & 2 & 2 &2\\
      Luminosity/IP   & ${\cal L}$     &  $10^{34} \mathrm {cm^2 s^{-1}}$ & 1.35 & 1.5 & 3 & 8.5\\
      Beam polarisation $e^- / e^+ $& ${\cal P}$& \% &80/30 & 80/0& 0/0 & 0/0\\ 
      Repetition frequency       & $f_{rep}$  & Hz & 5 & 50 & -- & --\\
      Bunches/pulse      & $n_{bunch}$  & 1 & 1312 & 352 & -- & --\\
      Bunches/beam      & $n_{beam}$  & 1 & -- & -- & 242 & 328\\
      Bunch population & $N_e$ & $10^{10}$ & 2 & 0.4 & 15 & 18\\
      Bunch interval   &  $\Delta t_b$ & ns & 554 & 0.5 & 680 & 827\\ 
      Bunch current/pulse   &  $I_{pulse}$ & mA & 5.8 & 2002 & -- & --\\
      Beam pulse duration  &  $t_{pulse}$ & $\mu$s & 727 & 0.176 & -- & --\\
      Beam current    &  $I_{beam}$ & mA & -- & -- & 17.4 & 29\\
      Normalised hor. emitt. at IP &  $\gamma \epsilon_x$ & $\mu$m & 5 & 0.95 & 284 & 148\\
      Normalised vert. emitt. at IP &  $\gamma \epsilon_y$ & nm & 35 & 30 & 727 & 305\\
      RMS hor. beam size &  $\sigma^{\ast}_x$ & nm & 516 & 149 & 20900 & 13748\\
      RMS vert. beam size &  $\sigma^{\ast}_y$ & nm & 7.7 & 2.9 & 68 & 36\\
      Average beam power/beam  &  $P_{beam}$ & MW & 2.65 & 3.35 & -- & --\\
      SR power/beam & $P_{SR}$ & MW & -- & -- & 30 & 50\\
      Site AC Power & $P_{site}$& MW & 127 & 252 & 270 & 308\\
      Site length &$L_{site}$& km& 20.5 & 22.8 & -- & --\\
      Ring circumference & $L_{ring}$ & km & -- & -- & 100 & 97.8\\ 
  \bottomrule
   \end{tabular}
   \caption{Parameters of the different proposals for energy frontier $e^+ e^-$ colliders (Status: December 2018). The choice has been oriented to those variants that are likely to become the respectively first stage. Disclaimer: The numbers have been carefully compiled or estimated from Refs.~\cite{Evans:2017rvt, bib:schulteclic-hh2018, bib:koratz-ias2018, bib:papa-hh2018, bib:gao-hh2018, Bordry:2018gri, CEPCStudyGroup:2018rmc}. All mistakes are mine and R.P. invites projects to point out mistakes or to add missing numbers.}
   \label{tab:eeparams}
\end{table}

\end{appendices}
\end{document}